\documentclass[aps,pra,preprint,groupedaddress]{revtex4-1}
\usepackage{graphicx}
\usepackage{setspace}
\usepackage{epsfig}
\usepackage{amsmath}
\usepackage{mathrsfs}
\usepackage{braket}
\usepackage{amssymb} 
\usepackage{mathtools}
\usepackage{graphicx}
\usepackage{textcomp}
\usepackage{amssymb}
\usepackage{slashbox}
\usepackage{geometry}
\usepackage{subfigure}
\usepackage[normalem]{ulem}
\usepackage{color}
\usepackage{float}

\begin{document}
\title{Polarons Explain Luminescence Behavior of Colloidal Quantum Dots at Low Temperature}
\author{Meenakshi Khosla, Sravya Rao, and Shilpi Gupta}
\email{corresponding author: ShilpiG@iitk.ac.in}
\affiliation{Department of Electrical Engineering, Indian Institute of Technology Kanpur, Kanpur-208016, UP, India} 

\begin{abstract}
Luminescence properties of colloidal quantum dots have found applications in imaging, displays, light-emitting diodes and lasers, and single photon sources. Despite wide interest, several experimental observations in low-temperature photoluminescence of these quantum dots, such as the short lifetime on the scale of microseconds and a zero-longitudinal optical phonon line in spectrum, both attributed to a dark exciton in literature, remain unexplained by existing models. Here we propose a theoretical model including the effect of solid-state environment on luminescence. The model captures both coherent and incoherent interactions of band-edge exciton with phonon modes. Our model predicts formation of dressed states by coupling of the exciton with a confined acoustic phonon mode, and explains the short lifetime and the presence of the zero-longitudinal optical phonon line in the spectrum. Accounting for the interaction of the exciton with bulk phonon modes, the model also explains the experimentally observed temperature-dependence of the photoluminescence decay dynamics and temperature-dependence of the photoluminescence spectrum.
\end{abstract}

\maketitle

\section*{Introduction}
Colloidal quantum dots show size-tunable luminescence spectrum \cite{Alivisatos1996, Klimov2003}, have high luminescence efficiency at room temperature \cite{Lee2000, Qu2002}, and can be easily functionalized for incorporation into a variety of systems \cite{DeMelloDonega2011}. These properties make them useful as biomarkers \cite{BruchezJr.1998}, gain materials for light emitting diodes \cite{Mashford2013, Qian2011} and lasers \cite{Min2006, Eisler2002}, and electroluminescent materials for displays \cite{Coe2002}. They are also potential candidates for single photon sources \cite{Brokmann2004a}.  
It is crucial to understand the fundamental mechanism for luminescence in these quantum dots to tailor their properties for different applications and to overcome issues like blinking \cite{Nirmal1996}, trapping of carriers \cite{Jones2003}, and non-radiative recombination processes \cite{Klimov2000} that hinder many potential applications. For these reasons, colloidal quantum dots have been the subject of many experimental \cite{Klimov2007b, Califano2005, Crooker2003, Biadala2009, Brokmann2004b,Fushman2005,Rakher2010,Rakher2011,Di2012,Gesuele2012,Gupta2013, Oron2009, Eilers2014, Biadala2016, Robel2015, Talapin2003, Talapin2004,Makarov2014,Choi2012, Kagan1996} and theoretical \cite{Efros1996, Leung1998, Franceschetti1999, Gupta2014, Rodina2015, Rodina2016, Manjavacas2011} studies.


Luminescence spectrum and decay dynamics of a quantum dot are governed by the band-edge exciton and its interaction with the solid-state environment. Numerous experiments and theoretical calculations have established that the fine structure of the band-edge exciton in a variety of colloidal quantum dots --- CdSe, InAs, CdTe, PbSe, ZnSe, InP/ZnS, and Ge --- includes the lowest-lying "dark" exciton state from which optical transition to ground state is forbidden, followed by a higher-energy "bright" exciton state from which optical transition to ground state is allowed \cite{Nirmal1996,Efros1996,Klimov2007b,Oron2009,Eilers2014,Biadala2016,Robel2015}. Because optical transitions are forbidden, a dark state should have an infinitely large radiative lifetime; in reality, interaction with solid-state environment reduces the lifetime. Even after taking such interactions into account, theoretical calculations predict dark exciton lifetimes on the order of milliseconds \cite{Califano2005,Califano2007}. However, experimentally observed lifetimes at $\sim$2 K, when most of the excited-state population is expected to reside in the lowest-lying dark exciton state, are found to be surprisingly lower, in the range of 0.35 $\mu$s - 81 $\mu$s \cite{Crooker2003,Oron2009,Eilers2014,Biadala2016,Robel2015}. 

For temperatures less than 20 K, photoluminescence decay of colloidal quantum dots exhibits two distinct lifetimes \cite{Crooker2003, Labeau2003}. At $\sim$2 K, the longer of these approaches the above mentioned microsecond-scale lifetime (the other being much smaller), and has been found to vary with temperature in a variety of colloidal quantum dots \cite{Crooker2003, Biadala2009,Brovelli2011, Raino2011,Labeau2003, DeMelloDonega2006,Oron2009,Werschler2016}. To explain the temperature dependencene, a three-level model with ground, dark, and bright states, and with thermal distribution of population between the dark and the bright states, has been widely used \cite{Labeau2003,Crooker2003,DeMelloDonega2006,Oron2009,Biadala2009,Brovelli2011,Eilers2014,Biadala2016,Werschler2016}. Although the model predicts the temperature-dependence of the lifetime, it does not explain the shortening of the lifetime of the dark exciton from theoretically predicted milliseconds to experimentally observed microseconds \cite{Califano2005,Biadala2009}. The model also does not explain other experimental observations. In the photoluminescence spectrum at 2 K, three closely and almost equally spaced peaks have been observed \cite{Biadala2009}. The lowest energy peak has been attributed to confined acoustic phonon sideband of the dark exciton, and the two higher energy peaks are attributed to zero-longitudinal phonon (LO) lines of the dark and the bright exciton. The existing theoretical model does not explain the existence of the lowest energy peak, and though the model assigns one of the peaks in photoluminecsence spectrum to zero-LO phonon line of the dark exciton, the origin of radiaitve decay of dark exciton is not understood \cite{Biadala2009}. In experiments, the relative strengths of the three peaks change and their linewidths increase with increase in temperature \cite{Biadala2009}.

To explain the short lifetime of the dark exciton at low temperatures, three hypotheses have been proposed. The first hypothesis attributed the shortening of lifetime to surface states which could result in mixing of the dark and the bright exciton states \cite{Califano2005}. However, this hypothesis is disproved by the experiments in which changes in surface passivation coating did not affect the lifetime \cite{Crooker2003, DeMelloDonega2006}. The second hypothesis attributes the shortening of lifetime to LO phonon-assisted recombination of the dark exciton \cite{Crooker2003}. Indeed, LO phonon sideband of the dark exciton has been observed in photoluminescence measurements \cite{Biadala2009}. However, experimental observation of a zero-LO-phonon line for the dark exciton in photlouminescence spectrum  \cite{Biadala2009} shows that the dark exciton must also have a direct radiative decay channel, unaccounted in the hypothesis. The third hypothesis states that the shortening of lifetime is caused by the exciton coupling to confined acoustic phonon modes \cite{Oron2009, Huxter2010, Eilers2014, Rodina2015, Rodina2016}; however no specific coupling mechanism has been proposed or demonstrated.

Here we propose a model to explain the unexpectedly short exciton lifetime in colloidal quantum dots at low temperatures, building upon a set of recent observations. Confined acoustic phonon modes have been observed in colloidal quantum dots in a variety of experiments \cite{Saviot1996,Woggon1996, Sagar2008, Oron2009, Chilla2008,GranadosDelAguila2014,Werschler2016}. Because the observed confined acoustic phonons have energy in the range of 1-10 meV, which matches with the observed energy gap between the bright and the dark excitons \cite{Efros1996, Crooker2003}, and because the phonon modes exhibit a characteristic discrete spectrum (long coherence) \cite{Chilla2008,Werschler2016}, strong exciton-acoustic phonon coupling in colloidal quantum dots is expected. This behavior is analogous to that seen in self-assembled indium arsenide quantum dots in which both experiments and theoretical calculations have shown formation of polarons by strong coupling of exciton and LO phonons; this coupling occurs because LO phonon energy often matches the energy gap between exciton states and because LO phonons have weak dispersion \cite{Preisler2006, Hameau2002, Hameau1999, Sarkar2005, Knipp1997, Verzelen2002}. Based on these observations, we propose that the photoluminescence spectrum and photoluminescence decay dynamics of colloidal quantum dots at low temperatures can be explained by strong coupling between exciton and a confined acoustic phonon mode, resulting in dressed exciton-phonon (polaron) states. 

Our model for the colloidal quantum dot at low temperature consists of a ground state, a dark exciton state, and a bright exciton state (as predicted from fine structure of band-edge exciton), where the excitonic states are coupled via a confined (coherent) acoustic phonon mode and also interact with a bath of bulk (incoherent and dispersive) phonons. The model predicts that the strongly-coupled exciton-confined acoustic phonon system has three excited energy eigenstates: dressed bright, dressed dark, and bare dark exciton states. Our calculations reveal that (a) the microsecond-scale excitonic lifetime observed at low temperatures is the radiative lifetime of the dressed dark state. The model also explains (b) the temperature-dependence of photoluminescence decay \cite{Crooker2003, Biadala2009,Brovelli2011, Raino2011,Labeau2003, DeMelloDonega2006,Oron2009,Werschler2016}, (c) the existence of the lowest energy peak in the three-peak spectrum and the radiative decay channel of the zero-LO phonon line of the dark exciton \cite{Biadala2009}, and (d) the temperature-dependence of the photoluminescence spectrum \cite{Biadala2009}.

\section*{Quantum Dot Model}
\begin{figure}[htbp]
\centering\includegraphics[width=\textwidth]{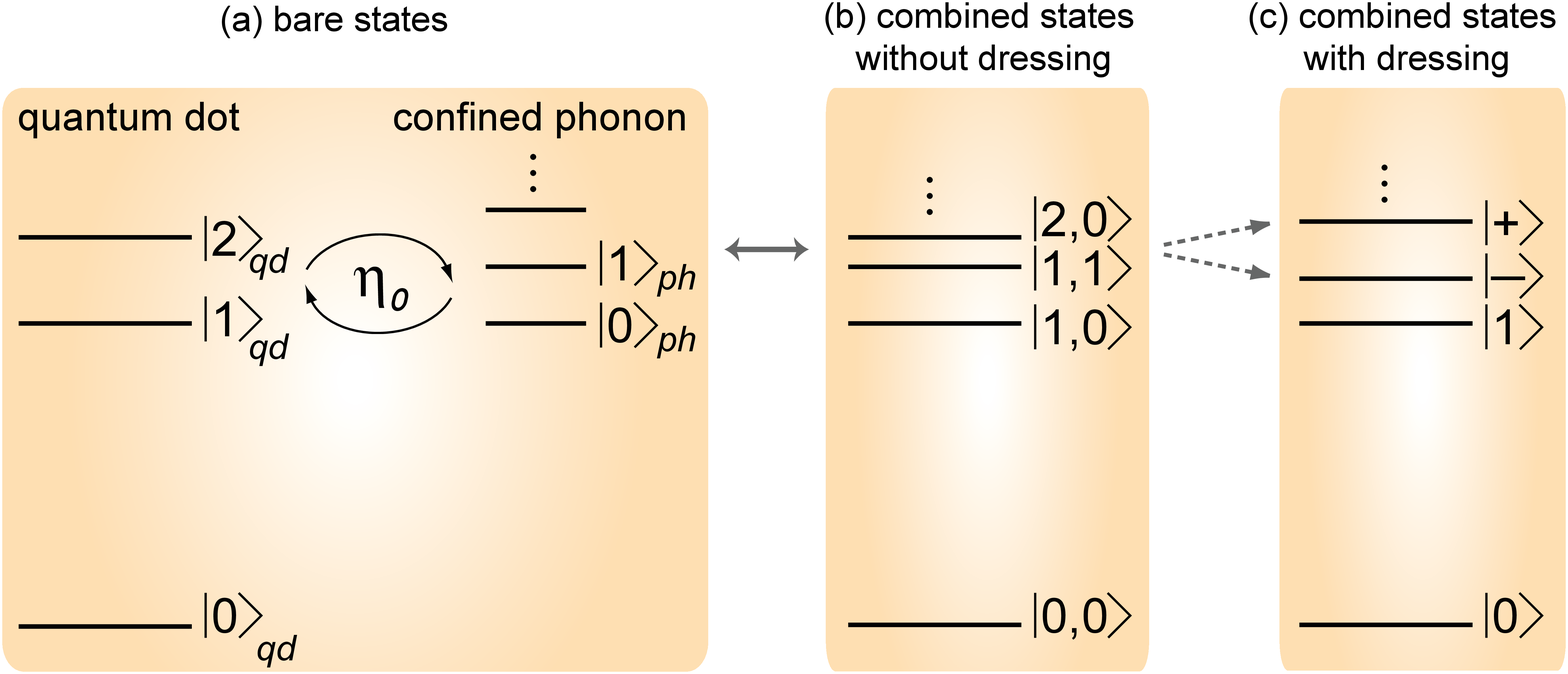}
\caption{Energy-level diagram of (a) a colloidal quantum dot interacting with a confined acoustic phonon mode in the bare-sate basis, (b) combined states of the exciton-phonon system without dressing (factorized states; notation: $\ket{i,j}$ where index $i \in [0,1,2]$ for the quantum dot states and index $j \in [0,1]$ for confined acoustic phonon states), and (c) combined states of the exciton-phonon system with dressing (dressed states).}
\label{Fig:BareQD_schematic}
\end{figure}
We model a colloidal quantum dot as a three level system consisting of a ground $\ket{0}_{qd}$, a dark exciton $\ket{1}_{qd}$ and a bright exciton $\ket{2}_{qd}$ state (Fig. \ref{Fig:BareQD_schematic}a), which is sufficient for modeling the photoluminescence behavior \cite{Klimov2007b}. The Hamiltonian accounting for the interaction of the quantum dot with a confined acoustic phonon mode, under rotating wave approximation, is
\begin{eqnarray}
\begin{aligned}
\label{eq:H_S}
\mathbf{H_{S}} &= \hbar\omega_{1}\sigma_{11}+ \hbar\omega_{2}\sigma_{22}+\hbar\omega_{ph}\mathbf{b_{0}^\dagger}\mathbf{b_{0}}+\hbar\eta_0(\sigma_{21}\mathbf{b_{0}}+\sigma_{12}\mathbf{b_{0}^\dagger)}\\
\end{aligned}
\end{eqnarray}
Here, we set the energy of the ground state of the quantum dot to zero. We define $\omega_{1}$ and $\omega_{2}$ as the resonant frequencies of the bare dark and the bare bright exciton states, and $\omega_{ph}$ as the frequency of the confined acoustic phonon mode (Fig. \ref{Fig:BareQD_schematic}a). The operator $\sigma_{jk} = |j\rangle\langle k|$ represents atomic dipole operator when $j \neq k$ and atomic population operator when $j = k$. Boson annihilation (creation) operator for confined acoustic phonon mode is $\mathbf{b_0}$ ($\mathbf{b_0}^\dagger$). The excitons interact with the confined acoustic phonon mode via deformation potential coupling \cite{Takagahara1996}, which is governed by $\eta_0$ (Fig. \ref{Fig:BareQD_schematic}a). Many experimental findings in colloidal quantum dots at low temperatures suggest that $\eta_0$ is greater than the decay rates of excitons and the confined acoustic phonon mode \cite{Krauss1997,Oron2009, Sagar2008, Califano2005, Labeau2003,Spann2013}. Therefore, we consider the excitonic states and the confined acoustic phonon mode to be strongly coupled. Diagonalization of system Hamiltonian, $\mathbf{H_S}$ (Eq. \ref{eq:H_S}), gives three excited eigenstates (dressed or polaron states) for the first ladder of the manifold \cite{Blais2004}: $\,\,\ket{+} = cos\theta\ket{2,0}-sin\theta\ket{1,1}$; $\,\,\ket{-} = cos\theta\ket{1,1}+sin\theta\ket{2,0}$; and $\,\,\ket{1} = \ket{1,0}$.

The corresponding eigen-frequencies are $\omega_{+}$, $\omega_{-}$, and $\omega_{1}$, where
\begin{equation}
\label{eq:eigenenergies_+-}
\omega_{\pm} = \frac{\omega_1 + \omega_2 + \omega_{ph}}{2} \pm \frac{\Delta}{2}\sqrt{1+\frac{4\eta_0^2}{\Delta^2}}
\end{equation}
Here $\theta=1/2 tan^{-1}(-2\eta_0/\Delta)$, $\Delta=\omega_2-\omega_1-\omega_{ph}$ is the detuning between the dark-bright splitting and the confined acoustic phonon mode energy, and $\ket{i,j}$ represents combined state of exciton-phonon system without dressing (factorized state) with index $i \in [0,1,2]$ for the quantum dot states and index $j \in [0,1]$ for confined acoustic phonon states (Fig. \ref{Fig:BareQD_schematic}b). Since we are interested in understanding luminescence behavior of the quantum dots at low temperature (upto 20 K), regime in which average phonon number is less than 1, we restrict our analysis to the first ladder of the manifold.

The radiative decay rates for the dressed states (Fig. \ref{Fig:BareQD_schematic}c) are 
\begin{eqnarray}
\begin{aligned}
\label{eq:DecayRate_DressedState2_+}
\Gamma_+  \approx \Gamma_2 cos^2\theta \\
\label{eq:DecayRate_DressedState2_-}
\Gamma_- \approx \Gamma_2 sin^2\theta 
\end{aligned}
\end{eqnarray}
where $\Gamma_2$ is the radiative decay rate of bare bright exciton state $\ket{2}_{qd}$; see Methods Eq.  \ref{eq:DecayRate_DressedState1_+-} for derivation.

To analyze the dynamics of the dressed system by including incoherent processes resulting from interaction with phonon bath and photon bath, we use master equation: $\frac{d}{dt} \rho =-\frac{i}{\hbar}[\mathbf{\bar{H}_S}, \rho] + \mathbf{L_{spont}}\rho + \mathbf{L_{ph}}\rho + \mathbf{L_{pump}}\rho$, where $\rho$ is the combined density matrix of the quantum dot-phonon system, $\mathbf{\bar{H}_S}$ is the system Hamiltonian in the dressed state basis (Methods Eq. \ref{eq:H_DressedState}), $\mathbf{L_{spont}}$, $\mathbf{L_{ph}}$, and $\mathbf{L_{pump}}$ represent Lindblad superoperators accounting for spontaneous emission, phonon scattering and pure dephasing, and incoherent pumping, respectively (Methods Eq. \ref{eq:L_ph}). We neglect terms containing bilinear functions of phonon operators in our analysis. Since $\Delta$ and $\eta_0$ are dependent on material and size of the quantum dot \cite{Takagahara1996,Sagar2008,Krauss1997}, a range of values are possible for both the parameters. Putting a conservative bound on the possible detuning in the system, we assume $\Delta > \eta_0$ . Under this condition, the dressed dark state (lower polaron, $\ket{-}$) is predominantly a phononic state and the dressed bright state (upper polaron, $\ket{+}$) is predominantly an excitonic state. Hence, we neglect interactions of $\ket{-}$ with excitonic states, $\ket{+}$ and $\ket{1}$, via a phonon.

\begin{figure}[htbp]
\centering\includegraphics[width=8 cm]{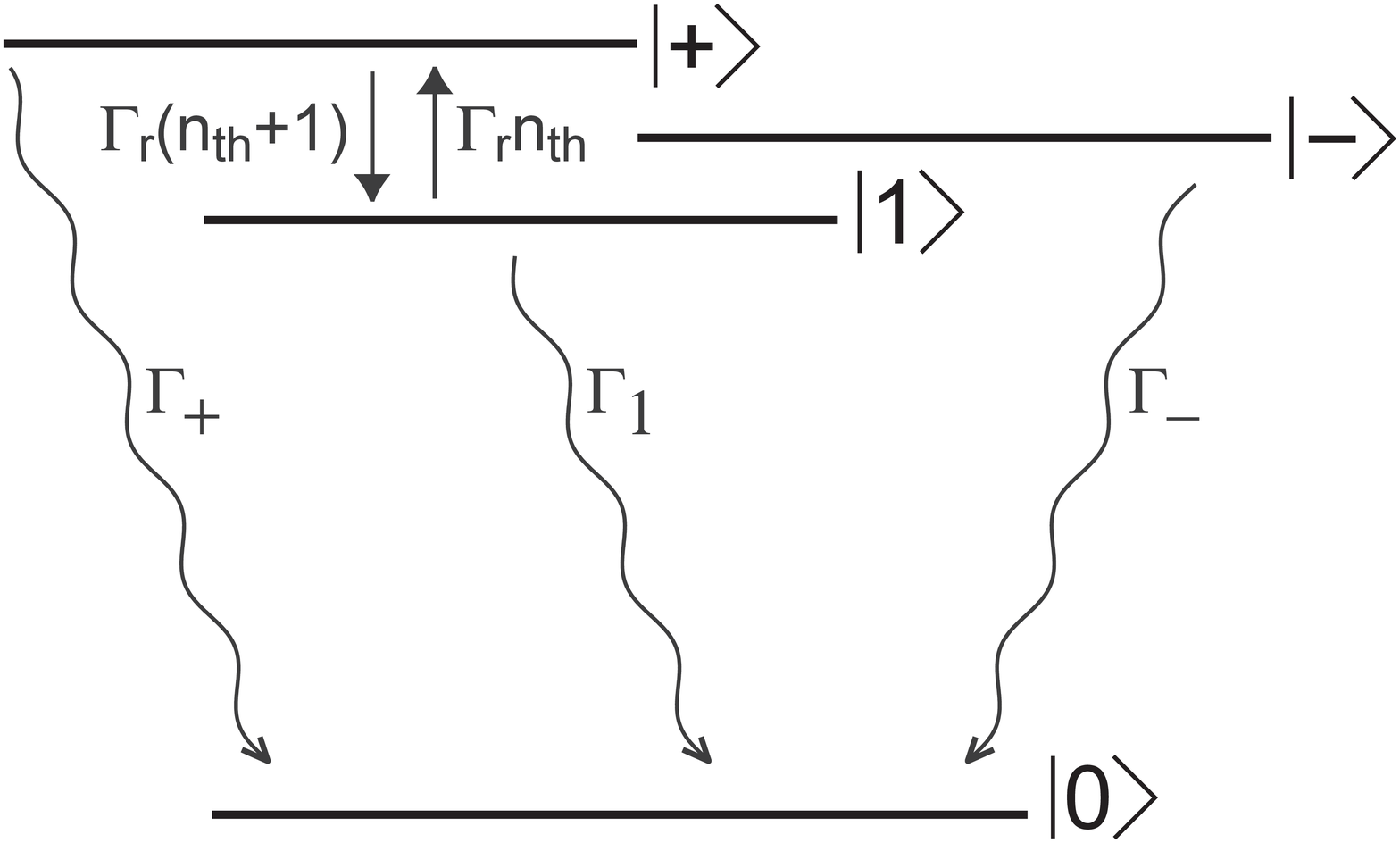}
\caption{Energy-level diagram of the dressed exciton-phonon system interacting with a photon bath and a phonon bath, showing spontaneous emission decay rates and spin-flip decay rates of excited energy-levels.}
\label{Fig:DressedState_schematic}
\end{figure}

We derive the equations of motion for the state-populations from the master equation as
\begin{equation}
\label{eq:EquationofMotion_population}
\begin{aligned}
\frac{d}{dt} 
\begin{bmatrix}
\rho_{++} \\
\rho_{--} \\
\rho_{11}
\end{bmatrix} = \begin{bmatrix}
-(\Gamma_{r}(n_{th}+1)+\Gamma_{+}+\Gamma_{P+}) & -\Gamma_{P+} & \Gamma_r n_{th}-\Gamma_{P+} \\
-\Gamma_{P-} & -(\Gamma_{-}+\Gamma_{P-}) & -\Gamma_{P-} \\
\Gamma_{r}(n_{th}+1)-\Gamma_{P1} & -\Gamma_{P1} & -(\Gamma_r n_{th}+\Gamma_1+\Gamma_{P1}) 
\end{bmatrix}
\begin{bmatrix}
\rho_{++} \\
\rho_{--} \\
\rho_{11}
\end{bmatrix}+\begin{bmatrix}
\Gamma_{P+} \\
\Gamma_{P-} \\
\Gamma_{P1} \\
\end{bmatrix}
\end{aligned}
\end{equation} 

and relevant dipole transition dynamics as

\begin{equation} 
\label{eq:EquationofMotion_dipole}
\begin{aligned}
\frac{d\rho_{0+}}{dt} &= i\omega_{+}\rho_{0+}-\kappa_{+}\rho_{0+}\\
\frac{d\rho_{0-}}{dt} &= i\omega_{-}\rho_{0-}-\kappa_{-}\rho_{0-}\\
\frac{d\rho_{01}}{dt} &= i\omega_{1}\rho_{01}-\kappa_{1}\rho_{01}\\
\end{aligned}
\end{equation}
where $\Gamma_r$ is bulk phonon-assisted spin-flip rate between the dressed bright $\ket{+}$ and the bare dark exciton $\ket{1}$ states, $\Gamma_{Pi}$ where $i\in[+,-,1]$, is incoherent pumping rate for different energy eigenstates, $\Gamma_1$ is the radiative decay rate of bare dark exciton state $\ket{1}$, and $n_{th} = (\exp{(\epsilon/kT) - 1})^{-1}$ is bulk phonon number under thermal equilibrium at temperature $T$ and energy $\epsilon = \hbar(\omega_+ - \omega_{1})$. Figure \ref{Fig:DressedState_schematic} shows the various energy levels and the population decay rates in the dressed state basis. The decay rates of the dipole moments (Eq. \ref{eq:EquationofMotion_dipole}) in the dressed state basis are $\kappa_{+} = (\Gamma_{+}+\Gamma_r(n_{th}+1)+\Gamma_\phi+\Gamma_P)/2$, $\kappa_{-} = (\Gamma_{-}+\Gamma_P)/2$, and $\kappa_{1} = (\Gamma_{1}+\Gamma_rn_{th}+\Gamma_\phi+\Gamma_P)/2$, where $\Gamma_P = \Gamma_{P+}+\Gamma_{P-}+\Gamma_{P1}$ and $\Gamma_\phi$ is pure dephasing rate. Using Eqs. \ref{eq:EquationofMotion_population}-\ref{eq:EquationofMotion_dipole} we next calculate expressions for steady-state luminescence spectrum and luminescence decay.

\section*{Steady-state Luminescence Spectrum}
To calculate expression for steady-state luminescence spectrum of this quantum dot-phonon system, we express the electric field operator at the detector \cite{Madsen2013} as $\mathbf{E^-}=\sqrt{\Gamma_+}\sigma_{0+} + \sqrt{\Gamma_-}\sigma_{0-} + \sqrt{\Gamma_1}\sigma_{01}$; because the separation between transition frequencies is small (1-10 meV), we assume the collection efficiency of the detector is the same for various transition frequencies. Using equations of motion of transition dipoles (Eq. \ref{eq:EquationofMotion_dipole}) and Quantum Regression Theorem \cite{Scully1997}, we calculate two-time correlation function of the electric field at the detector, and calculate steady-state luminescence spectrum, $S(\omega)$, using Wiener-Khinchin theorem \cite{Scully1997}  (Methods Eq. \ref{eq:WeinerKhinchin}). 
\begin{equation}
\label{eq:Spectrum}
S(\omega)=\frac{1}{\pi}\bigg(\frac{\Gamma_+\kappa_+\rho_{++,ss}}{(\omega-\omega_+)^2+\kappa_+^2} + \frac{\Gamma_-\kappa_-\rho_{--,ss}}{(\omega-\omega_-)^2+\kappa_-^2} + \frac{\Gamma_1\kappa_1\rho_{11,ss}}{(\omega-\omega_1)^2+\kappa_1^2}\bigg)
\end{equation}
The relative contributions of various transitions to the steady-state luminescence spectrum in Eq. \ref{eq:Spectrum} depend on respective steady-state populations of each excited state, $\rho_{ii,ss}$, where $i\in[+,-,1]$. We evaluate $\rho_{ii,ss}$ by solving Eq. \ref{eq:EquationofMotion_population} in steady state. Eq. \ref{eq:Spectrum} immediately provides an explanation for the observation \cite{Biadala2009} that the steady state spectrum consists of three peaks at eigenenergies (Eq. \ref{eq:eigenenergies_+-}) of the quantum dot-phonon system. All three peaks correspond to zero-LO phonon states. However, the middle peak --- assigned in literature to the zero-LO phonon line of the dark exciton ('F') \cite{Biadala2009} --- has been been a puzzling observation, as described above. Our model clears the confusion by qualifying the nature of the state further --- the middle peak corresponds to the zero-LO phonon line of the dressed dark state of the exciton-confined acoustic phonon system. The dressing imparts a partial bright character to the dark-exciton-one-phonon factorized state (Eq. \ref{eq:DecayRate_DressedState2_-}) due to which it appears in the emission spectrum. Similarly, our model shows that the highest energy peak that has been assigned in literature to the zero-LO phonon line of the bright exciton ('A') \cite{Biadala2009} is actually zero-LO phonon line of the dressed bright state of the exciton-confined acoustic phonon system. Further, the lowest energy peak has also been observed in experiments, and was assigned to the confined acoustic phonon sideband of the dark exciton ('Ac') \cite{Biadala2009}. However, this assignment has been problematic, as it also predicts a confined acoustic phonon sideband of the bright exciton, which has not been observed in experiments. Our model shows that the lowest energy peak actually corresponds to the zero-confined acoustic phonon line of the bare dark exciton state. 

\section*{Luminescence Decay Dynamics}
Next, we investigate the dynamics of luminescence decay of the quantum dot. Upon pulsed excitation at time $t =$ 0, the luminescence signal received by the detector \cite{Labeau2003, Biadala2016} is $I(t) = \rho_{++}(t)\Gamma_+ + \rho_{--}(t)\Gamma_- + \rho_{11}(t)\Gamma_1$, again assuming the collection efficiency of the detector is the same for all three transition frequencies. To calculate time-dependent expressions for populations, we solve coupled equations of motion, Eq. \ref{eq:EquationofMotion_population}, with pumping rates $\Gamma_{Pi} =$ 0, in terms of initial conditions $\rho_{ii}(0)$ where $i\in[+,-,1]$. The eigenvalues of the matrix in Eq. \ref{eq:EquationofMotion_population} represent eigen decay rates of the system, and are given by Eq. \ref{eq:EigenValues_DecayRate} in Methods section. In colloidal quantum dots, typical values for phonon assisted spin-flip times are $\sim$100 ps, bright exciton decay lifetime $\sim$10 ns, and expected dark exciton lifetime $\sim$1 ms \cite{Labeau2003,Biadala2009, Califano2005,Masia2012}. Therefore, $\Gamma_r >> \Gamma_+, \Gamma_1$; and $\Gamma_+ >> \Gamma_1$; these conditions simplify Eq. \ref{eq:EigenValues_DecayRate} to
\begin{eqnarray}
\begin{aligned}
\label{eq:EigenValues_DecayRate_simplified}
\alpha &=\Gamma_- \\
\beta &= \Gamma_r(2n_{th}+1)\\
\zeta &= \Gamma_+\frac{n_{th}}{2n_{th}+1} + \Gamma_1\frac{n_{th}+1}{2n_{th}+1}
\end{aligned}
\end{eqnarray}
Here, eigen decay rate $\alpha$ represents decay rate of the dressed dark state that is predominantly a phononic state. The population of the dressed dark state decays only due to spontaneous emission of photons, resulting in temperature-independent decay of its population. The other two states, the dressed bright and the bare dark exciton, exchange population via bulk phonons; their decay is characterized by a fast rate $\beta$ that represents the non-radiative decay of population due to a bulk phonon-assisted spin-flip process and a slow rate $\zeta$ that represents the radiative decay of populations due to spontaneous emission of photons. The signal received by the detector takes the form
\begin{equation}
\label{eq:Intensity_versus_time}
I(t)= I_{\alpha}e^{-\alpha t} + I_{\beta}e^{-\beta t} + I_{\zeta}e^{-\zeta t}
\end{equation}
where $I_{\alpha}$, $I_{\beta}$, and $I_{\zeta}$ represent contribution of respective eigen decay rates to the detected signal $I(t)$. These contributions depend on various decay rates of the system and initial conditions $\rho_{ii}(0)$ for $i\in[+,-,1]$ (Methods Eqs. \ref{eq:Populations_versus_time}-\ref{eq:IntensityCoefficients}). Eqs. \ref{eq:EigenValues_DecayRate_simplified}-\ref{eq:Intensity_versus_time} immediately provide an explanation for the temperature-dependent decay curves observed experimentally \cite{Crooker2003, Biadala2009,Brovelli2011, Raino2011,Labeau2003, DeMelloDonega2006,Oron2009,Werschler2016}. In the next section, we calculate and analyze photoluminescence spectrum and decay curves for a specific case of quantum dots.

\section*{Calculations for CdSe Quantum Dots} 
To perform calculations, we consider the specific case of CdSe colloidal quantum dots\cite{Crooker2003,DeMelloDonega2006,Biadala2009}. In time-resolved photoluminescence experiments on these quantum dots at 2 K, two distinct decays have been observed: an extremely fast decay and a slow decay with lifetime of $\sim$ 1 $\mu$s that is temperature independent below 2 K \cite{Crooker2003,DeMelloDonega2006,Biadala2009}. From Eq. \ref{eq:EigenValues_DecayRate_simplified}, at 2 K when $n_{th} \ll 1$, we associate the fast decay to $\beta$, and the slow and temperature-independent decay to $\alpha$. Therefore, we assign the observed 1 $\mu$s lifetime to the dressed dark state of our model and set $\Gamma_- =$ $10^{-3}$ ns$^{-1}$. The bright exciton is expected to have an intrinsic radiative lifetime of about 10 ns \cite{Califano2005, Labeau2003}, and therefore we set decay rate of bare bright exciton $\Gamma_2 =$ 0.1 ns$^{-1}$. Using Eq. \ref{eq:DecayRate_DressedState2_-}, and the values of $\Gamma_-$ and $\Gamma_2$, we estimate $\eta_0/\Delta =$ 0.1; this low value is not surprising when the system has not been engineered for maximising the coupling $\eta_0$ or minimizing the detuning $\Delta$. The above analysis also gives $\Gamma_+ =$ 0.1 ns$^{-1}$ (Eq. \ref{eq:DecayRate_DressedState2_+}). Since the bare dark exciton is expected to have lifetime in milliseconds \cite{Califano2007}, we set the decay rate of the bare dark exciton to $\Gamma_1 =$ 10$^{-6}$ ns$^{-1}$. We set $\hbar\eta_0 =$ 0.08 meV (see Supporting Information Fig. S1 for results with other values of $\eta_0$), in agreement with experimentally estimated values of less than $0.2$ meV \cite{Sagar2008,Krauss1997}. We set $\hbar\omega_{1} =$ 2000 meV. Further, we set bulk phonon-assisted spin-flip rate $\Gamma_r =$ 10 ns$^{-1}$ and pure dephasing rate as a linear function of temperature $T$, $\Gamma_{\phi} = BT$ where $B =$ 0.6 ns$^{-1}$\cite{Biadala2009,Takagahara1996, Masia2012}. With these parameters, strong coupling condition --- $4\eta_0$ greater than the quantum dot decay rates $\Gamma_2, \Gamma_1, \Gamma_{\phi}, \Gamma_r(n_{th}+1)$ and the decay rate of the confined acoustic phonon mode which has been shown to be $\sim$ 10$^{-3}$ ns$^{-1}$ \cite{bron2013nonequilibrium, Spann2013} --- is also satisfied for $n_{th} \leq 1$.

\begin{figure}[htbp]
\centering\includegraphics[width=8cm]{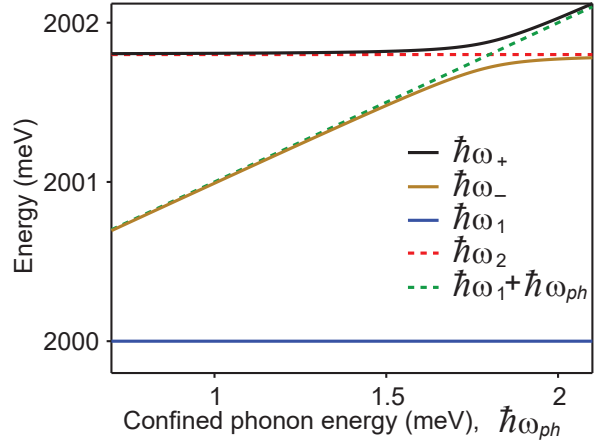}
\caption{Eigenenergies of the dressed states $\ket{+}, \ket{-}, \ket{1}$ and of the factorized exciton-phonon states $\ket{2,0}, \ket{1,1}, \ket{1,0} = \ket{1}$ as a function of the confined acoustic phonon mode energy.}
\label{Fig:Anti_Crossing}
\end{figure}

Figure \ref{Fig:Anti_Crossing} shows variation of eigenenergies of the dressed exciton-phonon system (shown by solid lines) and of the factorized exciton-phonon system (shown by dashed lines) with the energy of the confined acoustic phonon mode, in the absence of any incoherent and decay processes. The state with energy $\hbar\omega_{1}$ is common to both the dressed and the factorized exciton-phonon systems. The other two eigenenergies of the dressed system are $\hbar\omega_+$ and $\hbar\omega_-$, and they exhibit the signature anti-crossing behavior. The remaining exciton-phonon factorized states have energies $\hbar\omega_{2}$ and $\hbar(\omega_{1} + \omega_{ph})$. This figure shows that the dressed exciton-phonon system has three resonances in the emission spectrum. For further calculations, we set the confined phonon mode energy, $\hbar\omega_{ph} =$ 1 meV (see Supporting Information Fig. S1 for results with other values of $\hbar\omega_{ph}$).

\begin{figure}[htbp]
\includegraphics[width=\textwidth]{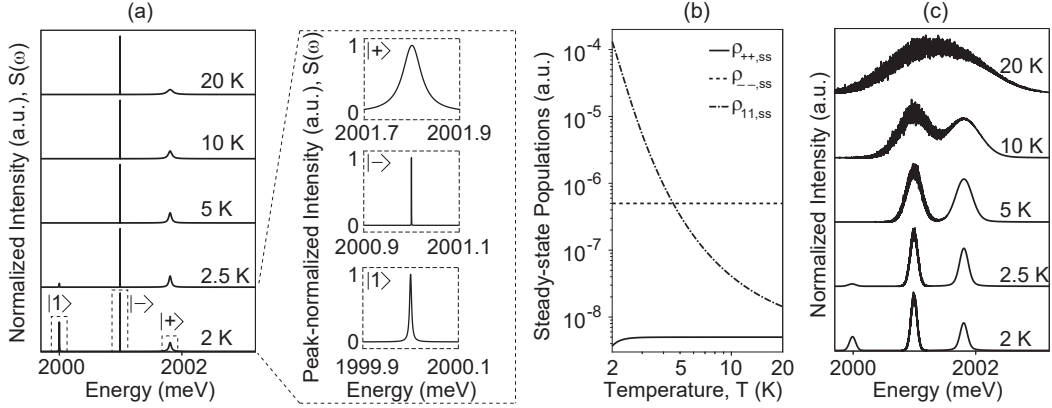}
\caption{(a) Normalized spectrum without spectral diffusion for different values of temperature; inset shows close-ups of normalized individual peaks at 2 K on the same energy scale. (b) Temperature variation of the steady state populations of the dressed states $\ket{+}, \ket{-}, \ket{1}$. (c) Normalized spectrum with spectral diffusion included for different values of temperature.}
\label{Fig:Spectrum}
\end{figure}

To estimate relative strength of the three peaks in the luminescence spectrum of the quantum dot and analyze how the spectrum changes with temperature, we calculate the steady state spectrum using Eq. \ref{eq:Spectrum}. We set equal pumping rate for the phononic state ($\ket{-}$) and the excitonic states combined (split equally in $\ket{+}$ and $\ket{1}$); $\Gamma_P =$ 10$^{-9}$ ns$^{-1}$ so that it is much smaller than all other decay rates of the system. Figure \ref{Fig:Spectrum}a shows that at T = 2 K, three peaks, as seen in experiments \cite{Biadala2009}, are present. 

The inset shows a closer view of the individual peaks at 2 K on a normalized intensity scale and the same energy scale. At 2 K, when $n_{th} \approx 10^{-5}$, the linewidth of the dressed bright state $\kappa_+ \approx \Gamma_r$, the linewidth of the dressed dark state $\kappa_- \approx \Gamma_{-}$ and the linewidth of the bare dark exciton state $\kappa_1 \approx \Gamma_{\phi}$ (Eq. \ref{eq:Spectrum}). Since $\Gamma_r$ is an order of magnitude larger than $\Gamma_{\phi}$ and four orders of magnitude larger than $\Gamma_{-}$, the linewidth of the dressed bright peak is much broader than the other two peaks. At low temperatures, the bulk phonon density, which is responsible for transfer of population between the bare dark exciton and dressed bright states, grows as $\exp{(-\epsilon/kT)}$ with temperature; therefore, as the temperature is increased, the population of bare dark exciton state rapidly decreases and of the dressed bright increases (Fig. \ref{Fig:Spectrum}b). This results in rapid decrease in intensity of the bare dark exciton peak in the spectrum (Fig.\ref{Fig:Spectrum}a). We calculate spectrum for a range of values of $\omega_{ph}$ and $\eta_0$ and observe similar behavior as in Fig. \ref{Fig:Spectrum}a (Supporting Information Fig. S1).

With increase in temperature, pure dephasing rate increases, which in turn increases the linewidth of the bare dark and the dressed bright exciton peaks (Fig. \ref{Fig:Spectrum}a). However, pure dephasing is not the dominant mechanism for observed linewidth broadening in colloidal quantum dots \cite{Empedocles1999}; in agreement with this, our calculations do not predict the broad linewidths (Fig.\ref{Fig:Spectrum}a) seen in experiments \cite{Biadala2009}. Rather, broadening of linewidths is primarily caused by spectral diffusion attributed to random fluctuation in the local environment of colloidal quantum dots \cite{Blanton1996,Empedocles1999,Neuhauser2000,Palinginis2003, Biadala2009,Sallen2010}. To incorporate the effect of spectral diffusion, we sample the transition frequencies from Gaussian distributions with means $\omega_1$ and $\omega_1 + \omega_{ph}$, as in Fig.\ref{Fig:Spectrum}a. Each step of spectrum computation, corresponding to one pair of frequencies sampled from the Gaussian distribution, is treated as 10-$\mu$s observation. The spectra are then integrated for a minute (i.e. 6 million steps). The standard deviations of the Gaussian distributions, reflecting the fluctuations in the local environment, are expected to depend on the temperature, although the precise relationship between spectral diffusion linewidth and temperature is unknown. We assume a simple, linear dependence on temperature: standard deviations is modeled as  6$T$ ns$^{-1}$, where the coefficient is arbitrarily chosen to resemble peak-width seen in experiments; we found that the observed behavior remains qualitatively unchanged if the coefficient (Supporting Information Fig. S2) or the nature of the temperature dependence is changed (Supporting Information Fig. S3).

Figure \ref{Fig:Spectrum}c plots steady state spectrum after 1-minute integration. As the temperature is increased, the linewidth of all the peaks increases and the peaks start to merge, a behavior that was also seen in experiments but was attributed to thermal mixing between the two higher energy states (dressed dark and dressed bright of our system) via acoustic phonons \cite{Biadala2009}. Our model rigorously accounts for temperature-dependence of quantum dot spectrum and provides an explanation for experimentally observed behavior. Note that, in the absence of spectral diffusion at 2 K, the dressed bright peak has a wider linewidth but lower intensity than the bare dark peak (inset of Fig. \ref{Fig:Spectrum}a). When spectra are integrated, summation of the tall-and-narrow bare dark peaks centered at slightly different frequencies results mostly in widening of the observed peak, while summation of short-and-wide dressed bright peaks results more in increasing the height of the observed peak. As a result, at 2 K, the intensity of the dressed bright peak is higher than that of the bare dark peak in the presence of spectral diffusion (Fig. \ref{Fig:Spectrum}c), although it was relatively lower in the absence of spectral diffusion (Fig. \ref{Fig:Spectrum}a).  

\begin{figure}[htbp]
\includegraphics[width=16 cm]{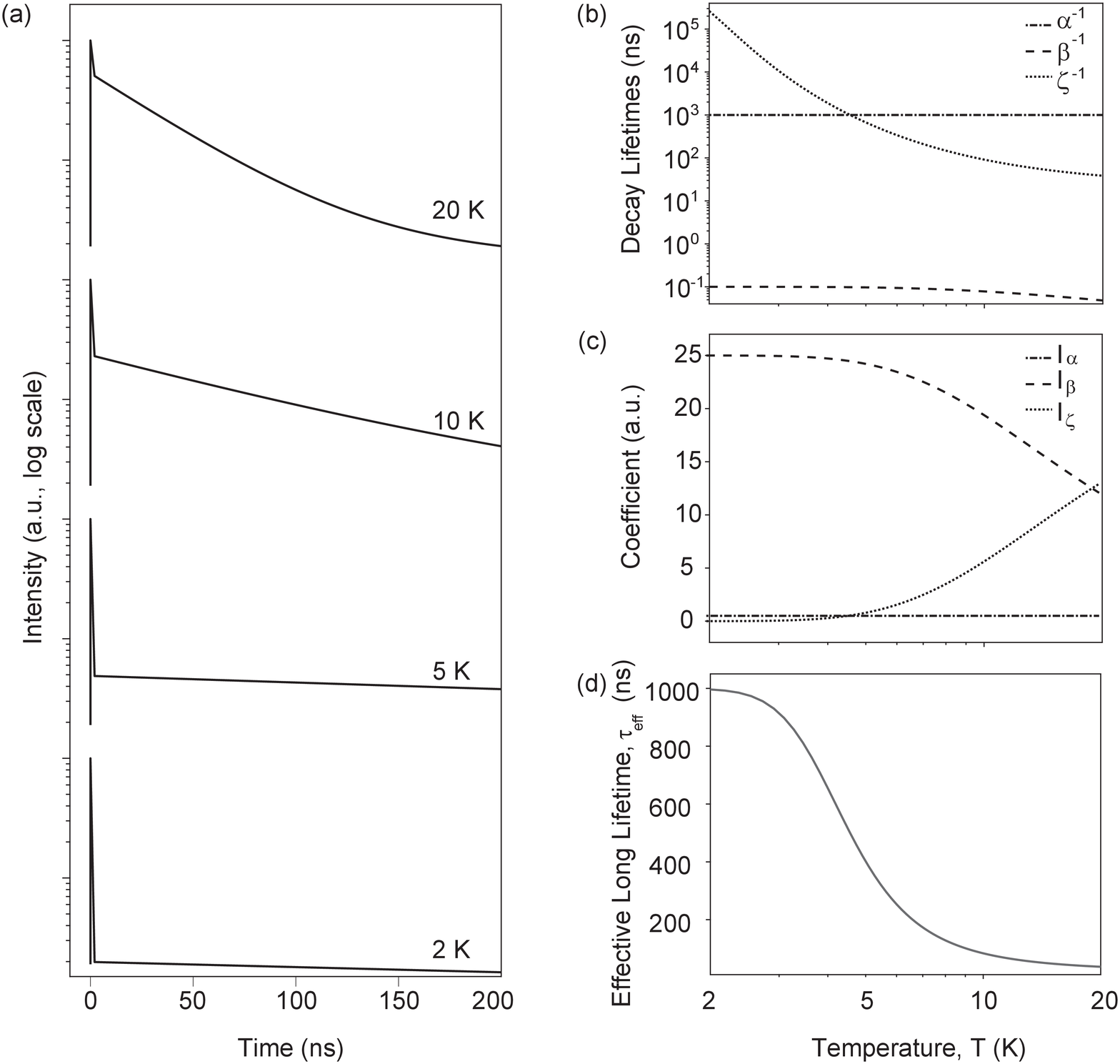}
\caption{(a) Decay curves for different values of temperature. (b) Eigenvalues of decay lifetimes as a function of temperature. (c) Temperature-dependence of contributions of different decay rates to the measured intensity. (d) Temperature-dependence of the effective long lifetime.}
\label{Fig:Lifetime}
\end{figure}

Next, we calculate the luminescence decay dynamics for specific case of CdSe quantum dots. To calculate time-dependence of luminescence intensity detected by the detector (Eq. \ref{eq:Intensity_versus_time}), initial populations $\rho_{ii}(0)$ where $i\in[+,-,1]$ are in the same proportion as pumping rates used while calculating spectrum, and coefficients $I_{\alpha}, I_{\beta}$, and $I_{\zeta}$ in Eq. \ref{eq:Intensity_versus_time} are evaluated using Eqs. \ref{eq:Populations_versus_time}-\ref{eq:IntensityCoefficients} from Methods section. Figure \ref{Fig:Lifetime}a plots quantum dot luminescence signal, $I(t)$ (Eq. \ref{eq:Intensity_versus_time}), as a function of time after pulsed excitation at time $t =$ 0, for different temperatures. At 2 K, when $n_{th} \ll$ 1, most of the population decays non-radiatively because of the fast phonon relaxation due to $\beta$ (Eq. \ref{eq:EigenValues_DecayRate_simplified}), and therefore, the intensity drops quickly initially and then it decays radiatively at a slow rate $\alpha$ (at 2 K, $\zeta \ll \alpha$, Fig. \ref{Fig:Lifetime}b). As temperature increases, radiative decay from the dressed bright state starts to compete against the phonon relaxation, which appears as a reduction in the initial drop and early onset of the slower decay of the photoluminescence intensity. This behavior is also seen in Fig. \ref{Fig:Lifetime}c that plots variation of coefficients $I_{\alpha}, I_{\beta}$, and $I_{\zeta}$ of Eq. \ref{eq:Intensity_versus_time} with temperature.  At 20 K, three decay rates are distinctly visible: the fastest decay is due to phonon-assisted non-radiative relaxation $\beta$; the next is $\zeta$ which has dominant contribution from radiative relaxation of dressed bright state ($\Gamma_+$); the slowest one is $\Gamma_-$ which is radiative relaxation of dressed dark state and is temperature-independent. The predictions of our model match well with the experiments \cite{DeMelloDonega2006}, and provide an explanation for the long standing question about the existence of microsecond-scale component of decay. Our calculations also explain (Fig. \ref{Fig:Lifetime}b) experimental observation of temperature-independent radiative lifetime below 2 K \cite{Crooker2003}. We calculate decay curves for another set of initial conditions, in which the bare dark exciton is not pumped and observe similar behavior as in Fig. \ref{Fig:Lifetime}a (Supporting Information Fig. S4), showing that the behavior of the model is not too sensitive to the initial conditions.

Many experimental investigations of decay dynamics of colloidal quantum dots have reported an S-shaped temperature dependence for the long-time component of decay \cite{DeMelloDonega2006, Biadala2009, Oron2009}. To further validate the findings of our model, we plot effective long lifetime, defined as $1/\tau_{eff} = \alpha + \zeta$ as a function of temperature in Fig. \ref{Fig:Lifetime}d, and observe an S-shaped dependence that matches well with the experimental data.  

\section*{Conclusions}
Multiple features observed in photoluminescence spectrum and photoluminescence decay of colloidal quantum dots at low temperatures ($<$ 20 K) have not been explained by a single model. These include: first, microsecond-scale lifetime attributed to the dark exciton; second, temperature-dependence of the photoluminescence decay; third, existence of the lowest energy peak in the three-peak spectrum and the radiative decay channel for the dark exciton that gives rise to the zero-LO phonon line in the spectrum; fourth, the temperature-dependence of the photoluminescence spectrum.

Our theoretical model provides a physical explanation for all four intriguing features observed in photoluminescence spectrum and photoluminescence decay. Based on experimental evidences from many works \cite{Saviot1996,Woggon1996, Sagar2008, Oron2009, Chilla2008,GranadosDelAguila2014,Werschler2016}, we propose that the excitonic states and a confined acoustic phonon mode are in strong coupling leading to formation of dressed dark and dressed bright states. Our calculations show that the microsecond-scale excitonic lifetime observed at low temperatures in experiments is the radiative lifetime of the dressed dark state. Our model also accounts for interaction of quantum dot with bulk phonon modes that impart temperature-dependent behavior. Our model provides a physical explanation for the S-shaped temperature dependence for the long-time component of decay. Further, the two dressed states along with the bare dark exciton state are the eigenenergy states of the coupled system, which appear in the photoluminescence spectrum at 2-3 K \cite{Biadala2009}. This explains the experimentally observed three-peaked photoluminescence spectrum \cite{Biadala2009}. Our model clarifies that the middle peak is the zero-LO phonon line of the dressed dark state of the strongly coupled exciton-phonon system, and therefore has a partial bright character that provides it a radiative decay channel. Finally, our model, with inclusion of spectral diffusion, closely predicts the temperature-dependence of the spectrum, including the observation that the peaks merge with each other around 20 K \cite{Biadala2009}.

Our model presents a significant advance over the existing and widely used theoretical model, which is based on thermal distribution of population via acoustic phonons between the dark and the bright exciton states \cite{Labeau2003,Crooker2003,DeMelloDonega2006,Oron2009,Biadala2009,Brovelli2011, Eilers2014,Biadala2016,Werschler2016}. This existing model does not explain the microsecond lifetime of the dark exciton, and the presence of the lowest energy peak in the three-peak spectrum and the origin of radiative decay channel for the dark exciton. The results from our model also indicate that the existing model does not characterize the middle peak sufficiently. While the middle peak is the zero-LO phonon line of the dark exciton, as characterized by the existing model, it is dressed dark state due to coupling between exciton and confined acoustic phonon mode, which our model predicts. 

In our model, we assumed initial conditions that the phononic state and the excitonic states combined are equally pumped to closely match the experimental results. However, changing the initials conditions will not affect the overall conclusions of our model. We assumed that spectral diffusion was linearly dependent on temperature; we also found that changing the relationship to Boltzmann distribution did not affect the conclusions. The precise relation is not known; however it is likely that diffusion will increase monotonically with temperature, and any such dependence is likely to give qualitatively similar results (for example, the merging of the dressed peaks at higher temperatures). The numbers used in our calculations were taken from CdSe quantum dots as these are the most studied; however, short exciton lifetime at low temperatures has been observed in a variety of colloidal quantum dots \cite{Crooker2003,Oron2009,Eilers2014,Biadala2016,Robel2015}.

In our model, we have not accounted for LO-phonon; it can easily be incorporated into this model as an additional, independent decay channel for the dark exciton (this should not affect other conclusions). The observed exciton lifetime at low temperatures has also been found to depend on size of the quantum dot; it can be incorporated in our model by including size-dependence of parameters like exciton-phonon coupling, exciton-phonon detuning, and intrinsic exciton decay rates. The model can be extended to higher temperatures by including the higher manifolds of the dressed states ladder. Our model provides a general framework on which these additional mechanisms can be easily added to provide a detailed description of the behaviors of colloidal quantum dots.

\section*{Methods} 
\subsection{Decay rate calculation in dressed-state basis} 
\label{appendix:DecayRate}
The transition dipole moment between the dressed states and the ground state $\ket{0,0}$ in the factorized exciton-phonon basis can be expressed as $\mu_{+,0}=\mu_{2,0}cos\theta-\mu_{1,1}sin\theta$ and $\mu_{-,0}=\mu_{2,0}sin\theta+\mu_{1,1}cos\theta $, where $\mu_{i,j} = \bra{0,0}\mu\ket{i,j}$ is the projection of dipole moment operator $\mu$ and $i\in[+,-,1,2]$ and $j =0,1$. 

The radiative decay rates for the dressed states in terms of transition dipole moment are expressed as \cite{Reynaud1977}
\begin{eqnarray}
\begin{aligned}
\label{eq:DecayRate_DressedState1_+-}
\Gamma_+ \propto |\mu_{+,0}|^2 = \mu_{2,0}^2 cos^2\theta+\mu_{1,1}^2 sin^2\theta-2\mu_{1,1}.\mu_{2,0}sin\theta cos\theta \\
\Gamma_- \propto |\mu_{-,0}|^2 =\mu_{2,0}^2 sin^2\theta+\mu_{1,1}^2 cos^2\theta+2\mu_{1,1}.\mu_{2,0}sin\theta cos\theta 
\end{aligned}
\end{eqnarray}
Since the bright state $\ket{2,0}$ is an optically active state, $|\mu_{2,0}| >> |\mu_{1,1}|$. This simplifies above expressions to Eq. \ref{eq:DecayRate_DressedState2_+}.

\subsection{Hamiltonian in dressed-state basis}
\label{appendix:Hamiltonian_Dressed}
The Hamiltonian representing bath and its interaction with quantum dot states
\begin{eqnarray}
\begin{aligned}
\label{eq:H_B}
\mathbf{H_B} &= \sum_q\hbar\omega_q \mathbf{b_q^\dagger}\mathbf{b_q} \\
\label{eq:H_SB}
\mathbf{H_{SB}} &= \sum_q \hbar\eta_{1q}\sigma_{11}(\mathbf{b_q}+\mathbf{b_q^\dagger})+\sum_q\hbar\eta_{2q} \sigma_{22}(\mathbf{b_q}+\mathbf{b_q^\dagger})\\
&+ \sum_q \hbar\eta_{12q}(\sigma_{21}\mathbf{b_q}+\sigma_{12}\mathbf{b_{q}^\dagger}) 
\end{aligned}
\end{eqnarray}
Here, $\mathbf{H_B}$ represents energy of the bulk phonon modes; and $\mathbf{H_{SB}}$ represents interaction of the quantum dot with bulk phonon modes. $\omega_{q}$ is frequency of the bulk phonon modes. Boson annihilation (creation) operator for bulk phonon modes is $\mathbf{b_q}$ ($\mathbf{b_q}^\dagger$). The bright and the dark exciton states of the quantum dot are coupled via bulk acoustic phonon modes with strength $\eta_{12q}$ that represents bulk phonon-assisted spin-flip process. The dark and the bright state also interact with the bath phonons with strengths $\eta_{1q}$ and $\eta_{2q}$, respectively, which account for pure dephasing mechanisms. To account for dephasing in a strongly coupled system, we need to consider the full Hamiltonian, including dephasing and scattering mechanisms, while deriving the master equation \cite{Carmichael1973, Pino2015}. 

The total Hamiltonian taking into account the interaction of quantum dot with both confined and bulk phonons, under rotating wave approximation, is $\mathbf{H}=\mathbf{H_{S}}+\mathbf{H_{B}}+\mathbf{H_{SB}}$. Neglecting terms with bilinear functions of phonon operators, and assuming $\eta_{1q} \approx \eta_{2q}=\eta_{q} $, i.e., dark and bright exciton states couple to the phonon modes with approximately the same coupling strength, the total Hamiltonian $H$ transforms to dressed-state basis as
\begin{equation}
\label{eq:H_DressedState}
\begin{aligned}
\mathbf{\bar{H}} &= \mathbf{\bar{H}_S} + \mathbf{\bar{H}_B} + \mathbf{\bar{H}_{SB}}\\
\mathbf{\bar{H}_S} &= \hbar\omega_{+}\sigma_{++} +\hbar\omega_{-}\sigma_{--}+\hbar\omega_{1}\sigma_{11}\\
\mathbf{\bar{H}_B} &= \sum_q \hbar\omega_q\mathbf{b_q^\dagger}\mathbf{b_q}\\
\mathbf{\bar{H}_{SB}} &= \sum_q \hbar\eta_q (\sigma_{++} +\sigma_{11})(\mathbf{b_q}+\mathbf{b_q^\dagger})\\
&+ \sum_q \hbar\eta_{12q} (\sigma_{+1}\mathbf{b_q}+\sigma_{1+}\mathbf{b_q^\dagger})
\end{aligned}
\end{equation}

\subsection{Linblad superoperators}
\label{appendix:Linblad}
\begin{eqnarray}
\begin{aligned}
\label{eq:L_spont}
\mathbf{L_{spont}}\rho &= \Gamma_+ L(\sigma_{0+})\rho+\Gamma_- L(\sigma_{0-})\rho+\Gamma_1 L(\sigma_{01})\rho\\
\label{eq:L_ph}
\mathbf{L_{ph}}\rho &= \Gamma_r(n_{th}+1) L(\sigma_{1+})\rho+\Gamma_r n_{th} L(\sigma_{+1})\rho\\
&+\Gamma_\phi L(\sigma_{++})\rho+\Gamma_\phi L(\sigma_{11})\rho\\
\label{eq:L_pump}
\mathbf{L_{pump}}\rho &= \Gamma_{P+}L(\sigma_{+0})\rho+\Gamma_{P-}L(\sigma_{-0})\rho+\Gamma_{P1}L(\sigma_{10})\rho
\end{aligned}
\end{eqnarray}
where Linblad superoperator is defines as $L(C)\rho = C\rho C^\dagger - \frac{1}{2}C^\dagger C\rho - \frac{1}{2}\rho C^\dagger C$ for collapse operator $C$. Bulk phonon scattering rate is $\Gamma_r$ and pure dephasing rate is $\Gamma_{\phi}$. Pumping rates for the dressed bright, dressed dark, and the bare dark exciton states are $\Gamma_{P+}$, $\Gamma_{P-}$, and $\Gamma_{P1}$. The system-bath Hamiltonian $\mathbf{\bar{H}_{SB}}$ in Eq. \ref{eq:H_DressedState} has a standard form of damped simple harmonic oscillator bath interaction that transforms to Linblad superoperator form of $\mathbf{L_{ph}}\rho$ in Eq. \ref{eq:L_ph} \cite{WallsMilburn2007,Majumdar2011}.

\subsection{Spectrum} 
\label{appendix:Spectrum}
Steady state spectrum expression from Wiener Khinchin theorem \cite{Scully1997}
\begin{equation}
\label{eq:WeinerKhinchin}
S(\omega)=\frac{1}{\pi}\lim_{t \to \infty} \text{Re} \int_0^\infty \! \left< \mathbf{E^{-}}(t)\mathbf{E^+}(t+\tau)\right>e^{i\omega\tau}\mathrm{d}\tau 
\end{equation}

\subsection{Decay dynamics} 
\label{appendix:DecayDynamics}
The eigenvalues of the matrix in Eq. \ref{eq:EquationofMotion_population} represent eigen decay rates, and are given by
\begin{eqnarray}
\begin{aligned}
\label{eq:EigenValues_DecayRate}
\alpha &=\Gamma_- \\
\beta, \zeta &=\frac{1}{2}(\Gamma_++\Gamma_1+\Gamma_r(2n_{th}+1))[1\pm\sqrt{1-\frac{4\{\Gamma_+\Gamma_1+\Gamma_r(\Gamma_+n_{th}+\Gamma_1(n_{th}+1))\}}{(\Gamma_++\Gamma_1+\Gamma_r(2n_{th}+1))^2}}]     \\
\end{aligned}
\end{eqnarray}

The time-dependent populations are
\begin{eqnarray}
\begin{aligned}
\label{eq:Populations_versus_time}
\rho_{++}(t) &= u_1e^{-\beta t} + u_2e^{-\zeta t}\\
\rho_{--}(t) &= \rho_{--}(0)e^{-\alpha t}\\
\rho_{11}(t) &= v_1e^{-\beta t} + v_2e^{-\zeta t}\\
\end{aligned}
\end{eqnarray}
where $u_1, u_2, v_1$, and $v_2$ are determined from the initial conditions $\rho_{ii}(0)$ where $i\in[+,-]$. The contributions of the three eigen decay rates to the signal received by the detector in Eq. \ref{eq:Intensity_versus_time} are
\begin{eqnarray}
\begin{aligned}
\label{eq:IntensityCoefficients}
I_{\alpha} &= \Gamma_-\rho_{--}(0)\\
I_{\beta} &= \Gamma_+u_1 + \Gamma_1v_1\\
I_{\zeta} &= \Gamma_+u_2 + \Gamma_1v_2\\
\end{aligned}
\end{eqnarray}

\section*{Acknowledgement}
We thank Harshawardhan Wanare and Saikat Ghosh for discussions and careful reading of the manuscript. SG acknowledges funding support from IITK (initiation grant) and SERB-Ramanujan fellowship (SB/S2/RJN-134/2014).

\bibliography{References}

\end{document}


\center{\textbf{Supporting Information}}\\
\center{\textbf{Polarons Explain Luminescence Behavior of Colloidal Quantum Dots at Low Temperature}}\\
\center{Meenakshi Khosla, Sravya Rao, and Shilpi Gupta}\\

\bigskip

\begin{figure}[htbp]
\renewcommand{\figurename}{\textbf{Figure S}}
\includegraphics[width=\textwidth]{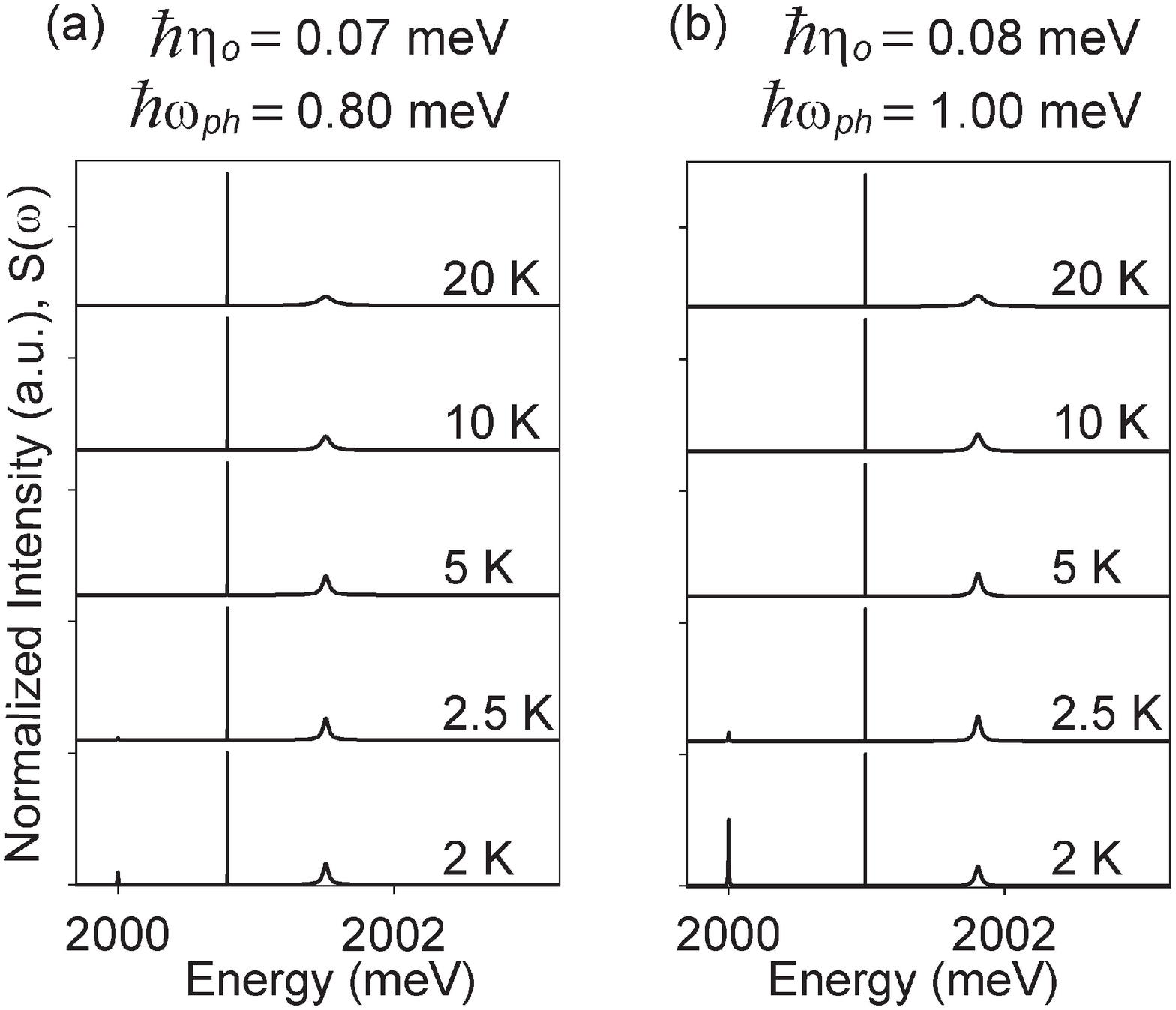}
\caption{Spectrum for different values of $\omega_{ph}$ and $\eta_0$. (a)$\hbar\eta_0 =$ 0.07 meV and $\hbar\omega_{ph} =$ 0.80 meV. (b)$\hbar\eta_0 =$ 0.08 meV and $\hbar\omega_{ph} =$ 1.00 meV.  (c)$\hbar\eta_0 =$ 0.09 meV and $\hbar\omega_{ph} =$ 1.20 meV. Spectral diffusion is not included.}
\label{Fig:Spectrum_extra}
\end{figure}

\begin{figure}[htbp]
\renewcommand{\figurename}{\textbf{Figure S}}
\includegraphics[width= 10 cm]{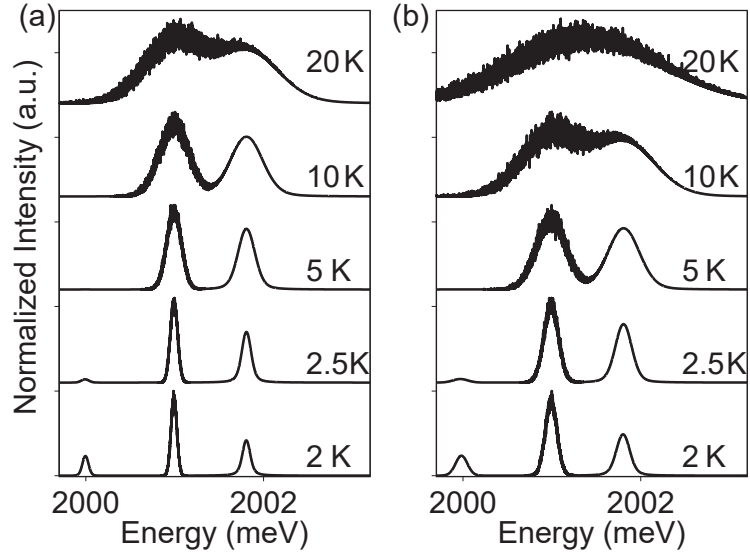}
\caption{Spectrum including spectral diffusion as a Gaussian distribution with standard deviation varying linearly with temperature $T$. Standard deviation (a) 4$T$ ns$^{-1}$. (b) 8$T$ ns$^{-1}$.}
\label{Fig:Spectrum_extra_SpectralDiffusion_Linear}
\end{figure}

\begin{figure}[htbp]
\renewcommand{\figurename}{\textbf{Figure S}}
\includegraphics[width=12 cm]{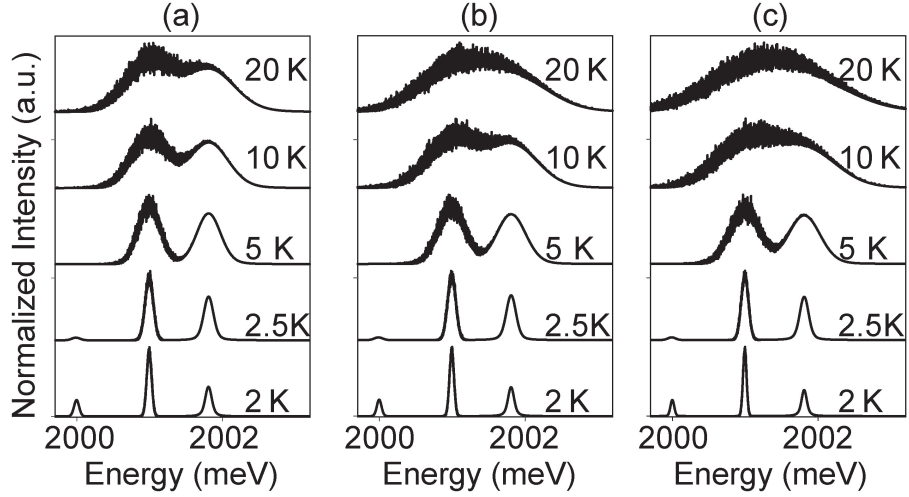}
\caption{Spectrum including spectral diffusion as a Gaussian distribution with standard deviation varying with temperature $T$ as Boltzmann Distribution. Standard deviation (a) 100e$^{-5/T}$ ns$^{-1}$. (b) 150e$^{-6/T}$ ns$^{-1}$.  (c) 200e$^{-7/T}$ ns$^{-1}$.}
\label{Fig:Spectrum_extra_SpectralDiffusion_exp}
\end{figure}

\begin{figure}[htbp]
\renewcommand{\figurename}{\textbf{Figure S}}
\includegraphics[width= 5cm]{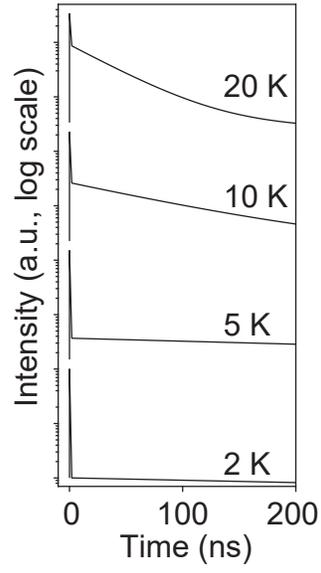}
\caption{Decay curves for a different set of initial conditions. $\rho_{++}(0) = 0.5, \rho_{--}(0) = 0.5, \rho_{11}(0) = 0$.}
\label{Fig:Lifetime_extra}
\end{figure}